\documentclass[journal=aamick,manuscript=article]{achemso}
\usepackage{lmodern}
\usepackage[bookmarks,%
            colorlinks,%
            urlcolor=blue,%
            citecolor=blue,%
            linkcolor=blue,%
            hyperfigures,%
            pdfcreator=KN,%
            breaklinks=false]{hyperref}
\usepackage[english]{babel}
\usepackage{siunitx}
\usepackage{csquotes}
\usepackage{nth}
\usepackage{multirow}
\usepackage[version=4]{mhchem}
\usepackage{graphicx}
\usepackage[capitalize]{cleveref}
\usepackage[caption=false]{subfig}

\DeclareGraphicsExtensions{.pdf}

\title{Enhancement of piezoelectric properties in a narrow cerium doping range of \ce{Ba_{1-x}Ca_{x}Ti_{1-y}Zr_{y}O3} evidenced by combinatorial experiment}
\author{Kevin Nadaud}
\affiliation[GREMAN]{%
GREMAN UMR 7347, Université de Tours, CNRS, INSA-CVL, 16 rue Pierre et Marie Curie, 37071 Tours, France%
}%
\email{kevin.nadaud@univ-tours.fr}
\author{Guillaume F. Nataf}
\affiliation[GREMAN]{%
GREMAN UMR 7347, Université de Tours, CNRS, INSA-CVL, 16 rue Pierre et Marie Curie, 37071 Tours, France%
}%
\author{Nazir Jaber}
\affiliation[GREMAN]{%
GREMAN UMR 7347, Université de Tours, CNRS, INSA-CVL, 16 rue Pierre et Marie Curie, 37071 Tours, France%
}%
\author{Béatrice Negulescu}
\affiliation[GREMAN]{%
GREMAN UMR 7347, Université de Tours, CNRS, INSA-CVL, 16 rue Pierre et Marie Curie, 37071 Tours, France%
}%
\author{Fabien Giovannelli}
\affiliation[GREMAN]{%
GREMAN UMR 7347, Université de Tours, CNRS, INSA-CVL, 16 rue Pierre et Marie Curie, 37071 Tours, France%
}%
\author{Pascal Andreazza}
\affiliation[ICMN]{%
ICMN, CNRS, Université d'Orléans, 1b rue de la Férollerie, CS 40059, 45071 Orléans Cedex 02, France%
}
\author{Pierre Birnal}
\affiliation[ICMN]{%
ICMN, CNRS, Université d'Orléans, 1b rue de la Férollerie, CS 40059, 45071 Orléans Cedex 02, France%
}
\author{Jérôme Wolfman}
\affiliation[GREMAN]{%
GREMAN UMR 7347, Université de Tours, CNRS, INSA-CVL, 16 rue Pierre et Marie Curie, 37071 Tours, France%
}%

\keywords{Combinatorial Pulsed Laser Deposition, Hyperbolic analysis, relaxor, domain wall, piezoelectricity, combinatorial experiment}

\begin{document}

\begin{abstract}
    Lead-free materials based on the \ce{(Ba,Ca)(Zr,Ti)O3} (BCZT) system exhibit excellent electromechanical properties that can be strongly modified by small amounts of dopants. 
    Here, we use a combinatorial strategy to unravel the influence of aliovalent doping with Ce on dielectric and piezoelectric properties of BCTZ. 
    We synthesize and characterize a single BCTZ thin film with a composition gradient from undoped to 0.2 mol\% cerium doping. 
    The cerium doping increases the piezoelectric coefficient from \qty{42.3(29)}{\pm\per\V} (undoped) to \qty{63(2.4)}{\pm\per\V} for 0.06 Ce-mol\%, and then decreases to \qty{38.4(13)}{\pm\per\V} for the maximum amount of cerium (0.2 mol\%). 
    An investigation of sub-coercive field non-linearities reveal that these variations are not only induced by changes in dynamics and densities of domain walls. 
    The results highlight the advantage of combinatorial techniques to identify ideal compositions for applications, without synthesizing a high number of samples with unavoidable sample-to-sample variations.
\end{abstract}
\maketitle

\section{Introduction}

Thanks to their high dielectric permittivity and high piezoelectric coefficient, relaxor ferroelectrics are promising material for energy storage and actuators\cite{JayakrishnanPMS2023,VeerapandiyanM2020}.
In this context, \ce{BaTiO3}-based materials such as the \ce{Ba_{1-x}Ca_{x}Ti_{1-y}Zr_{y}O3} (BCTZ) solid solution represent an interesting alternative to lead-based materials \cite{SimonJACLCOM2018,DaumontJAP2016,PuliJPD2019,YanJMC2020}.

For \ce{BaTiO3} derived materials, large piezoelectric coefficients have been found in the vicinity of the morphotropic phase boundary (MPB)\cite{AcostaAPR2017}.
It results from a complex interplay between chemistry, phase transitions and microstructures. For example, the dynamics of domain walls, in their ferroelectric phases, can enhance the electromechanical response\cite{ZhengAFM2023,CarpenterJPCM2015,DamjanovicJACS2005}.
In order to explore complex ternary diagrams using conventional synthesis, numerous samples are required, which can be challenging.
Furthermore, unavoidable sample-to-sample variations introduce uncertainties into measured properties.
In such context, combinatorial experiment emerged during the last decade to facilitate the exploration of phase diagrams by synthesizing samples with a continuous composition, structure or thickness variation \cite{LudwigNCM2019}.
For the elaboration of samples with such gradients, combinatorial pulsed laser deposition (CPLD) can be used \cite{LiuJAP2010,DaumontJAP2016,DaumontCoatings2021,Wolfmanbook2020}.
In addition to the application motivation, i.e. finding the optimum composition to maximize a given property, the combinatorial techniques permit to continuously explore phase diagrams and allows a fine determination of phase transitions.

In this article, we use the combinatorial experiment to investigate the effect of cerium doping on dielectric and piezoelectric properties of \ce{0.5(Ba_{0.7}Ca_{0.3}TiO3)-0.5(BaZr_{0.2}Ti_{0.8}O3)} thin films.
We chose to work with cerium since there exist already some reports on the influence of cerium doping on the dielectric and piezoelectric properties of BCTZ ceramics, revealing enhanced properties in a narrow composition range (0.025 to 0.25 mol\% of Ce) \cite{HayatiJAC2019} with few (3) tested compositions. 
The literature also contains contradictory results, e.g. on the evolution of phase transition temperatures with different Ce content \cite{CuiCI2012,ChandrakalaAIPConferenceProceedings2016,MaDJNB2021,ChandrakalaTICS2020}. 
Furthermore, the influence of Ce-doping on a BCTZ thin film has not been investigated so far.

Two samples were realized: an undoped BCTZ sample to calibrate the sensitivity of the piezoelectric characterization tool and to verify the homogeneity of the deposition technique and a second sample with a composition gradient from undoped to \qty{0.2}{mol\%} cerium doping.
In addition to conventional dielectric spectroscopy, sub-coercive field non-linearities have been studied using modified Rayleigh analysis and discussed in terms of modification of dielectric and piezoelectric properties.

\section{Experiments}

Polycristalline BCTZ film were deposited on \ce{Pt/TiO2 /MgO} substrates (\qtyproduct{10x10}{\mm}, (001) oriented) by combinatorial pulsed laser deposition (CPLD) technique \cite{LiuJAP2010,DaumontJAP2016} in a vacuum chamber with a base pressure of \qty{5e-8}{mbar}. 
A MgO substrate was preferred to Si to minimize thermal stress during the cooling process. 
A homogenized excimer laser beam (KrF, $\lambda=$\qty{248}{nm}, \qtyrange[range-phrase = --]{12}{20}{ns}) pass through a square aperture whose focused image is swept over the surface of a homemade \ce{Ba_{0.85}Ca_{0.15}Ti_{0.9}Zr_{0.1}O3} ceramic target (\ce{BCTZ50}) or a homemade \qty{0.2}{\%} Ce-doped \ce{Ba_{0.85}Ca_{0.148}Ce_{0.002}Ti_{0.9}Zr_{0.1}O3} ceramic target (\ce{Ce02{:}BCTZ50}). 
The resulting ablation plume scans over the entire substrate surface to ensure a good thickness and composition uniformity. 
Substrate temperature, oxygen pressure, laser fluence and repetition rate were optimized with respect to phase purity, crystallinity, surface roughness and piezoelectric performances. 
During the growth of the films, the laser fluence was kept at \qty{1.15}{\J\per\cm\squared}. 
The substrate temperature was held at \qty{700}{\celsius} with a dynamic \ce{O2} pressure of 0.3 mbar. 
Cooling was realized under 500 mbar static oxygen pressure at \qty{3}{\celsius\per\minute}  cooling rate, with a \qty{400}{\celsius} dwell step of one hour to ensure good oxygenation of the film.

\begin{figure}
    \centering
    \subfloat[]{%
    \resizebox{0.4\textwidth}{!}{%
    \includegraphics[width=\linewidth]{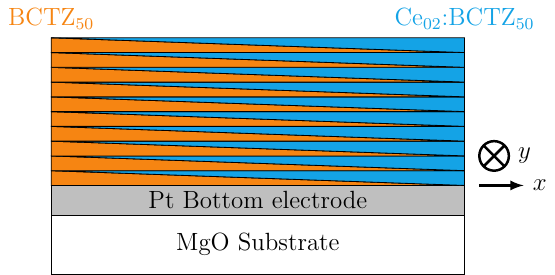}
    }%
    \label{subfig:sample side}}

    \subfloat[]{%
    \resizebox{0.2\textwidth}{!}{%
    \includegraphics[width=\linewidth]{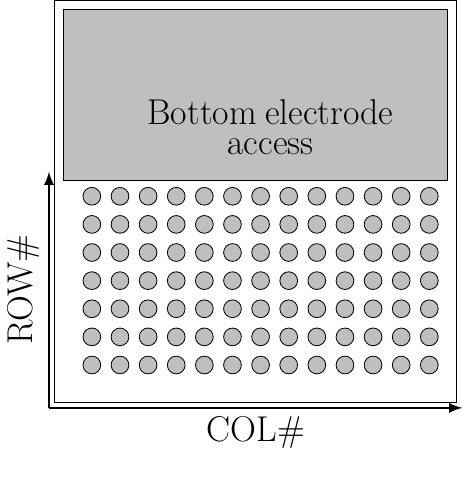}
    }%
    \label{subfig:sample picture}}
    \subfloat[]{%
    \resizebox{0.28\textwidth}{!}{%
    \includegraphics[width=\linewidth]{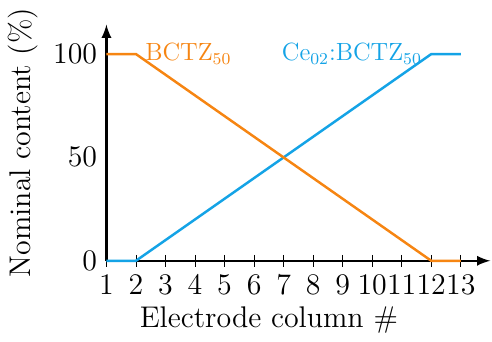}
    }%
    \label{subfig:sample gradient}}
    \caption{Side view schematic of the sample with a doping gradient (a). Schematic of the sample with row/column convention (b).  Nominal gradient for the \ce{Ce_{x}{:}BCTZ} sample (c).}
    \label{fig:picture gradient}
\end{figure}

Two samples were synthesized.
The first one is a reference sample deposited from a single undoped BCTZ target, (30000 laser pulses fired at \qty{6}{Hz}). 
The second one was deposited by combinatorial PLD using the undoped BCTZ target and a Ce-doped BCTZ target (\qty{0.2}{\% mol}). 
The combinatorial approach takes advantage of the very directional PLD plume to localize the deposition at the surface of the substrate using a mask with a rectangular aperture (shadow masking). 
Sliding this aperture, close to the substrate, in one direction combined with a small deposition rate (39 pulses / perovskite unit cells uc with the mask) allow \enquote{distributing} the deposited material on the surface.
By synchronizing the mask movements with the laser pulses fired on the undoped target, a controlled wedge of BCTZ is fabricated in a first deposition step: one side of the substrate \enquote{sees} 156 laser pulses, corresponding to 4 unit cells (uc), while on the opposite substrate side there is no deposition (see schematic \cref{subfig:sample side}).
Next, the complementary sequence of mask positioning and laser firing on the doped target is realized to complete the 4 uc thickness on the substrate surface with Ce-doped BCTZ.
This two-steps deposition sequence is repeated 250 times, targeting a final thickness of about \qty{400}{nm} (1000 uc).  
The resulting film has a composition that varies from \ce{BCTZ50} to \ce{Ce_{02}{:}BCTZ50} in the $x$ direction, while it is kept constant in the $y$ direction (Fig.~\ref{subfig:sample side}). 
Mask-substrate relative positioning at deposition temperature has an accuracy of a few hundreds of microns, leading to an offset of the composition gradient versus nominal position. 
To localize the experimental gradient and access its total nominal extent, \qty{1}{mm} wide bands with constant composition are framing the composition gradient in the sample design (Fig.~\ref{subfig:sample gradient}). 

Top Au electrodes (diameter \qty{330}{\micro\meter}) were then sputtered through a shadow mask.
An array of 91 capacitors is thus defined with 13 columns along the gradient direction and 7 lines for repetitions (Fig.~\ref{subfig:sample picture}) for the \ce{Ce_{x}{:}BCTZ50} doped sample.
With the lateral constant composition bands, 11 different compositions are accessible (columns 2 to 12). 
The same mask design was used for the undoped \ce{BCTZ50} sample, leading to the definition of 91 capacitors with the same nominal composition to characterize the deposition homogeneity.

Phase analysis and local structural investigations were conducted using X-ray micro-diffraction (parallel beam of Cu $K_{\alpha}$ mean radiation, Bruker Discover). 
Absolute global chemical composition and thickness were determined by Rutherford Back Scattering (RBS) on homogeneous \ce{BCTZ50} sample while local composition/thickness variations of both \ce{Ce_{x}{:}BCTZ50} and \ce{BCTZ50} were characterized by Wavelength Dispersive Spectroscopy  (WDS). 
Conventional WDS (or EDS) model assume a homogeneous sample at the electron probe level. 
The layered nature of our samples imposed to acquire emitted photons at various e-beam acceleration voltages. 
The voltage dependency of the intensity ratio coming from sample and standards (a.k.a. $K$-ratio) were then simulated by STRATAGEM software in order to extract thickness/composition variations for Ba, Ca, Ti and Zr (exemple shown Fig.~S3). 
The small Ce doping level (up to 0.2 mol\%) prevented Ce detection using WDS. 
Surface sensitive X-ray Photoelectron Spectroscopy (XPS) has been attempted to evidence the Ce doping gradient.   

The dielectric characterizations presented in this article have been acquired using a lock-in amplifier (MFLI with MD option, Zurich Instrument) connected to a temperature-controlled probe station, having a motorized chuck (Summit 12000, Cascade Microtech).
The AC measuring signal has been generated using the embedded lock-in generator.
Its amplitude has been swept from \qty{10}{mV_{rms}} to \qty{1}{V_{rms}} at a frequency of \qty{10}{kHz}.
The applied voltage and current through the capacitor are measured simultaneously by the lock-in and are demodulated simultaneously.
$|V|\exp\left(j\theta_{V}\right)$ is the phasor representing the applied voltage and $|I|\exp\left(j\theta_{I}\right)$ the phasor representing the current (with $j$ the imaginary unit).
The first harmonic of the current and the applied voltage are used to compute the complex impedance:\cite{NadaudJALCOM2022,NadaudAPL2021,NadaudJAP2023,NadaudAPL2024}
\begin{equation}
    Z = \frac{|V|}{|I|}\exp\left(j\left(\theta_{V}-\theta_{I}\right)\right)
\end{equation}
The complex capacitance $C^{*}$ can hence be derived from the complex impedance:
\begin{equation}
  C^{*} = \frac{1}{j\omega Z} = \frac{|I|}{\omega|V|}\exp\left(j\left(\theta_{I}-\theta_{V} - \frac{\pi}{2}\right)\right)
\end{equation}
with $\omega$ the angular frequency of the measuring voltage.
The material relative permittivity $\varepsilon_{r}^{*}$ is: 
\begin{equation}
  \varepsilon_{r}^{*} = \frac{t}{S\varepsilon_{0}}C^{*},
\end{equation}
with $t$ the thickness of the film, $S$ the surface of the electrodes and $\varepsilon_{0}$ the vacuum permittivity.
In the present case, the electrodes are sufficiently thick to limit the effect of the series resistance on the measured impedance.

For the phase transition temperature determination $T_{\mathit{max}}$, an AC measuring voltage of \qty{10}{mV_{rms}} has been preferred in order to avoid underestimations due to domain wall motion contribution\cite{NadaudAPL2024}.
$T_{\mathit{max}}$ have been determined from the least square fitting of $\varepsilon_{r}'$ with a parabola\cite{MaAPL2013,NadaudAPL2024}.

The ferroelectric and piezoelectric characterizations have been performed with an AixACCT DBLI and TF2000 ferroelectric analyzer (aixACCT Systems GmbH, Germany) using a triangular waveform at a frequency of \SI{1}{kHz} and a magnitude of \qty{5}{V}.
An automated chuck allows measuring the properties as function of the position of the electrode.
To ensure a good reflection of the laser beam, aluminum has been deposited on the back side of the sample.
The current is measured using the virtual ground method and the polarization is computed using numerical integration with respect to the time.
To improve the signal to noise ratio, the results of $P(E)$ and $S(E)$ loops correspond to the average over 1000 periods.
The calculation of the piezoelectric coefficients has been made using a linear regression from the maximum, respectively minimum field value to \SI{0}{\kV\per\cm} for positive, respectively negative, fields.
The linear regression has two advantages (i) it is less sensitive to noise and (ii) it provides a confidence indicator on the extracted value, coming from the regression algorithm. \cite{NadaudJALCOM2022}.
The presented piezoelectric coefficient is the average value between the piezoelectric coefficient for positive and for negative field.
To investigate the asymmetry of the $S(E)$ loop, the difference between positive and negative fields piezoelectric coefficient is also computed and noted $\Delta d_{33}^{*}$.

\section{Result and discussion}
The properties of undoped BCTZ sample are initially presented, focusing homogeneity of thickness, composition, dielectric and piezoelectric properties, as described in the first section of the article. 
Then, the dielectric and piezoelectric properties of the Ce-doped BCTZ sample (\ce{Ce_{x}{:}BCTZ50}) are presented and compared to the reference properties measured on the \ce{BCTZ50} sample.
In the third section, the effect of cerium doping on domain wall motion contribution is investigated.

\subsection{Undoped BCTZ}
X-ray diffraction pattern shows that the \ce{BCTZ50} film is single-phase to the detection limit and poly-oriented (see Fig. S1 in supplementary information of ref \cite{NadaudAPL2024}). 
Peak positions could be indexed according to orthorhombic \ce{BCTZ50} (icdd pdf \#04-022-8189). 
However the weak diffraction intensity from the poly-oriented film prevents to conclude on the exact symmetry (orthorhombic, rhombohedral, tetragonal,...). 
A diffraction pattern acquired at higher angle ($\qty{82.8}{\degree}< 2\theta < \qty{84.6}{\degree}$) (see Fig. S1 in supplementary material) exhibits an asymmetric peak with at least three contributions. 
This implies the presence of more than one symmetry as it is to be expected close to a MPB. 
The \ce{BCTZ50} film composition from RBS characterization (see Fig. S2a,b ref \cite{NadaudAPL2024}) was found to be close to the nominal target composition determined by WDS (see table S1 in supplementary material).

The local composition of the \ce{BCTZ50} sample has been studied using WDS (e-beam footprint of \qty{10}{\um}) and the results are reported as a function of the position on the sample (see Fig.~S3 and S4 in supplementary material).
For all elements, the ratio between $3\sigma$ dispersion and the mean value is below \qty{3}{\%}.

The thickness has been extracted as a function of the position using WDS (see Fig.~S4 in supplementary material).
An average thickness of \qty{370}{nm} is obtained for the undoped sample, neglecting its outskirt. 
When considering the whole sample, the ratio between $3\sigma$ dispersion and the mean value is \qty{10}{\%} but it decreases to \qty{7}{\%} when discarding edges of the sample.
All these data confirm the high uniformity of the composition and thickness of the \ce{BCTZ50} sample.

\begin{figure}
    \centering
    \subfloat{%
    \includegraphics[width=0.48\textwidth]{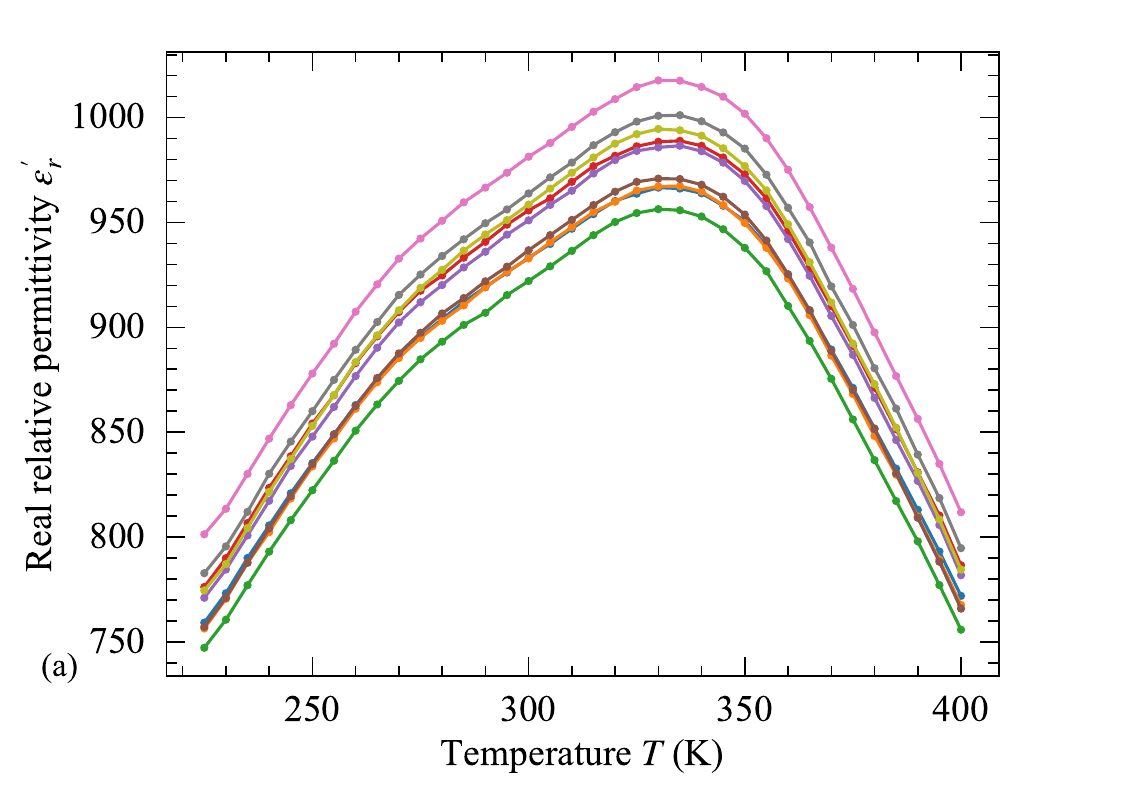}\label{subfig:NJ400 realperm TEMP}}\\%
    \subfloat{%
    \includegraphics[width=0.48\textwidth]{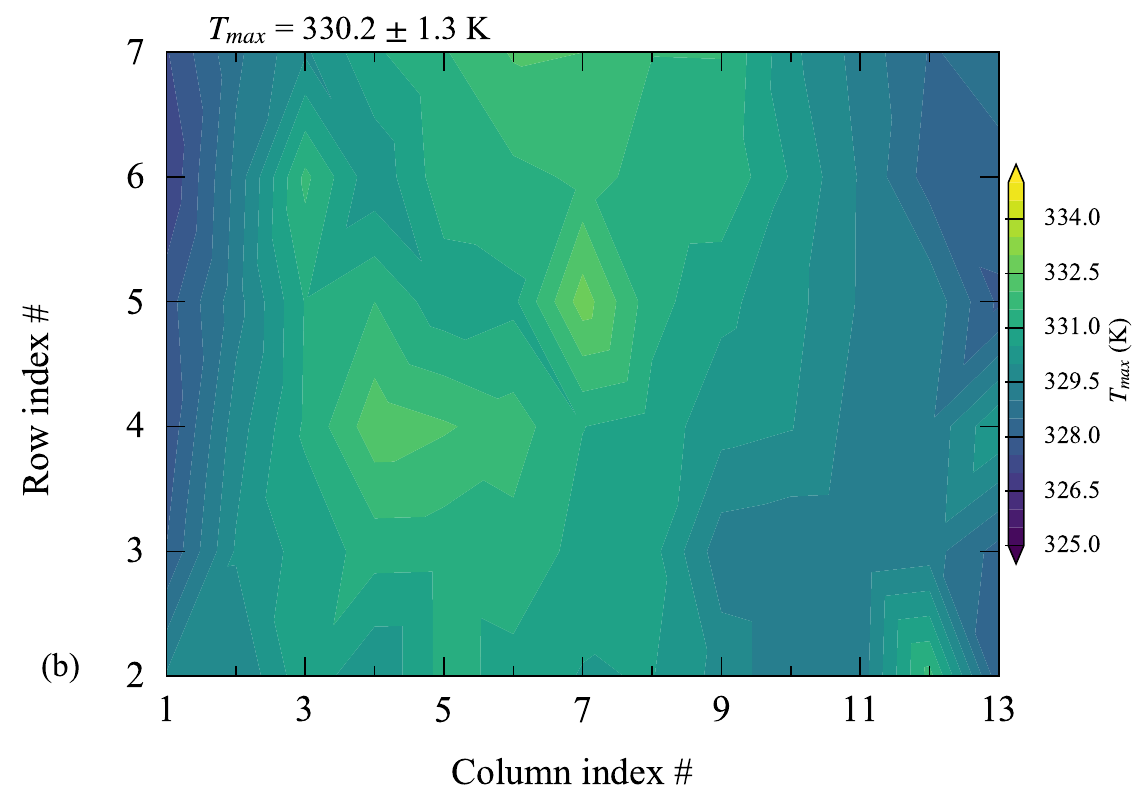}\label{subfig:NJ400 realPermTmax map}}%
    \caption{Relative permittivity as function of temperature for some selected capacitors (a) and $T_{\mathit{max}}$ map (b) for undoped \ce{BCTZ50} sample.}
    \label{fig:NJ400 realPerm}
\end{figure}

Fig.~\ref{subfig:NJ400 realperm TEMP} shows the dielectric permittivity as a function of temperature, for various representative electrodes on the sample.
The temperature evolution follows the same trend for all electrodes, with a maximum at \qty{330.2(13)}{K}, which is attributed to the ferroelectric to paraelectric phase transition\cite{NadaudAPL2024}.
This value is closed to what has been found for BCTZ thin films in literature (\qty{323}{K}) \cite{LiuCI2023,XuJACS2015}.
The transition temperature is highly homogeneous for all electrodes on the sample (Fig.~\ref{subfig:NJ400 realPermTmax map}) since a standard deviation of \qty{1.3}{K} is obtained.
A ferroelectric-ferroelectric transition is visible around \qty{270}{K} which may correspond to T-O transition observed in bulk \ce{BCTZ50}. 

\begin{figure*}
    \centering
    \includegraphics[width=0.48\textwidth]{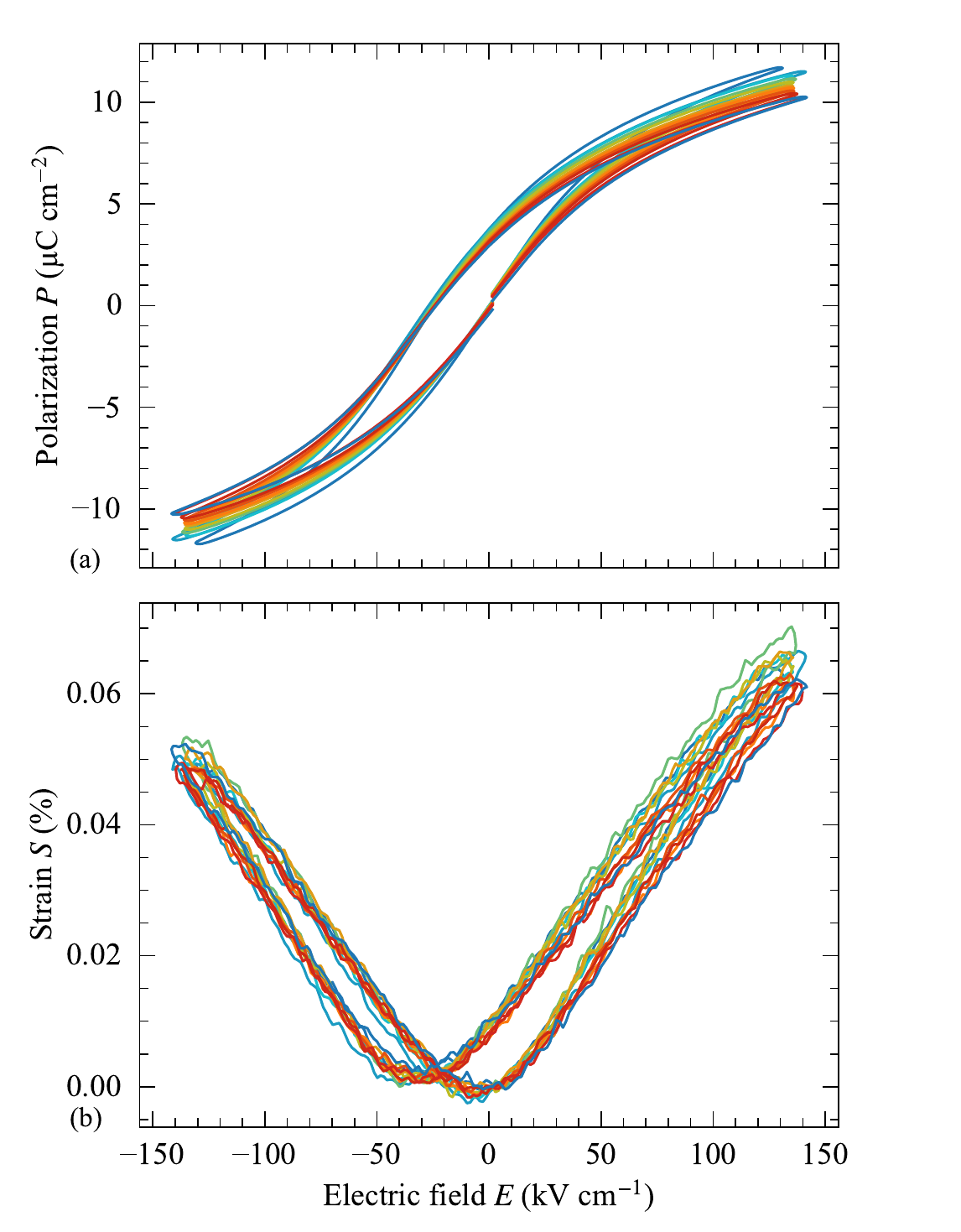}%
    \subfloat{\label{subfig:NJ400 polarization}}%
    \subfloat{\label{subfig:NJ400 strain}}%
    \includegraphics[width=0.48\textwidth]{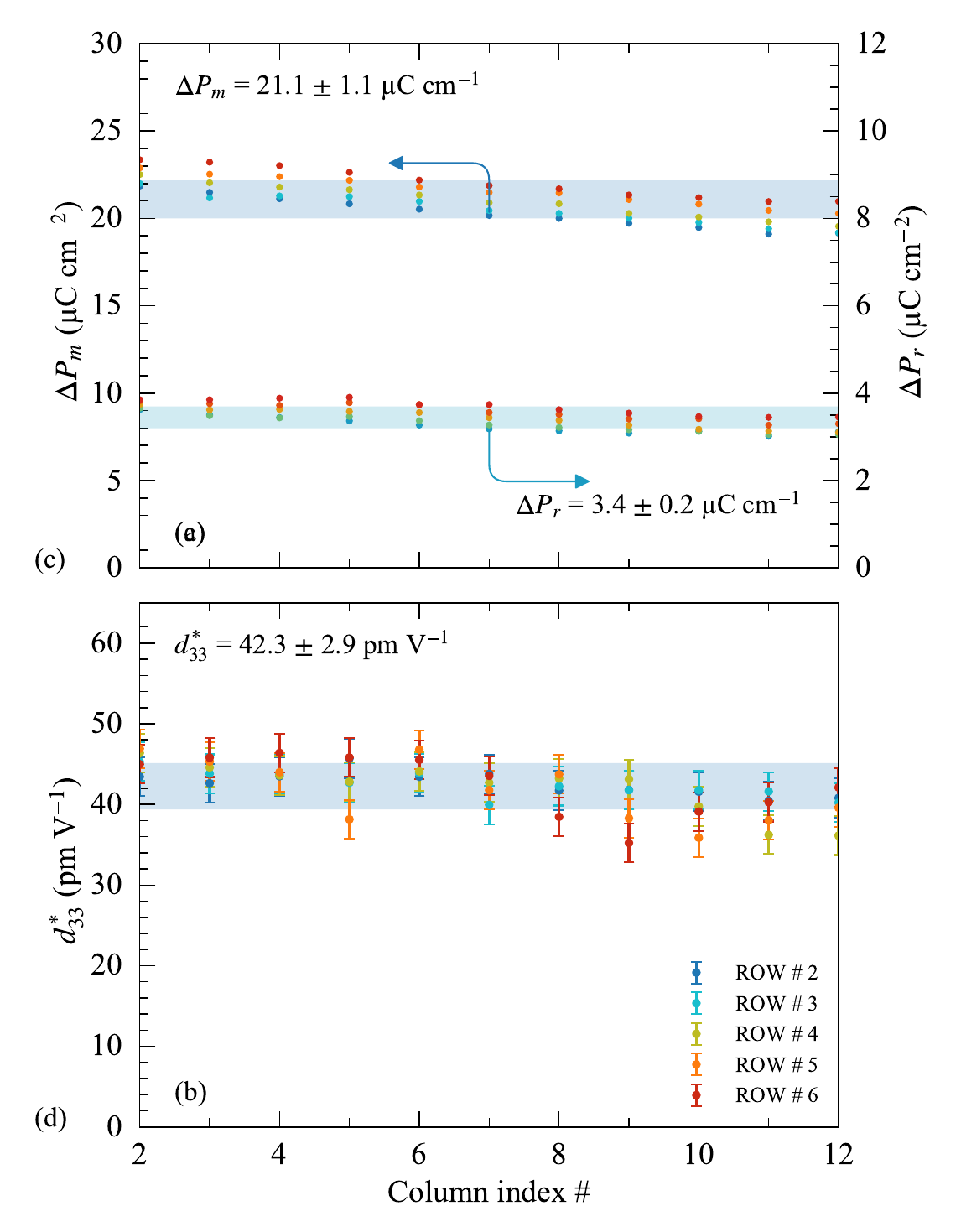}%
    \subfloat{\label{subfig:NJ400 2Pm2Pr data}}%
    \subfloat{\label{subfig:NJ400 d33 data}}%
    \caption{Polarization (a) and strain (b) as a function of the applied electric field for various electrodes on the \ce{BCTZ50} sample. Maximum polarization and remnant polarization (c) and piezoelectric coefficient (d) for the \ce{BCTZ50} sample. The shaded area represents the mean value and the standard deviation interval. The error bar for the $d_{33}^{*}$ data points corresponds to the measurement resolution, estimated from 10 successive measurements of a single capacitor.}
    \label{fig:NJ400 polarization strain}
\end{figure*}

Fig.~\ref{subfig:NJ400 polarization},\subref*{subfig:NJ400 strain} show the polarization and the strain for several electrodes on the sample.
The $P(E)$ loop is typical of relaxor ferroelectrics with a small remnant polarization and a slim shape.
Using the $P(E)$ loop, maximum and remnant polarizations have been computed for all capacitors and reported in Fig.~\ref{subfig:NJ400 2Pm2Pr data}.
Average values of $\Delta P_{m} = \qty{21.1(11)}{\micro\coulomb\per\cm\squared}$ and $\Delta P_{r} = \qty{3.4(2)}{\micro\coulomb\per\cm\squared}$ are obtained for the sample.

The horizontal shift of $P(E)$ and $S(E)$ loops is well visible since coercive field values are $E_{c}^{+} = \qty{2.5(0.9)}{\kV\per\cm}$ and $E_{c}^{-} = \qty{-33.1(1.6)}{\kV\per\cm}$.
This shift reveals the presence of an internal bias field, which can result from Schottky barriers at the electrodes and/or internal defects and polar nanoregions (PNRs) \cite{NadaudAPL2024}.

Fig.~\ref{subfig:NJ400 2Pm2Pr data},\subref*{subfig:NJ400 d33 data} presents the maximum polarization, the remnant polarization and the piezoelectric coefficient as a function of the position on the sample.
The average value of the piezoelectric coefficient is \qty{42.3(29)}{\pm\per\V}.
The DBLI resolution for piezoelectric coefficient determination was estimated from 10 successive measurements of a single capacitor, leading to a dispersion of \qty{2.4}{\pm\per\V} ($1\sigma$). 
This compares with the dispersion found across the sample (\qty{2.9}{\pm\per\V}).

All the results presented in this first part show the high uniformity of the composition and the dielectric/piezoelectric properties on a homogeneous \ce{BCTZ50} sample and establish an upper limit for the resolution in $d_{33}^{*}$ characterization. 

\subsection{Ce-doped BCTZ}
The composition of the \ce{Ce_{x}{:}BCTZ50} sample has been measured using the same procedure as described previously and the results are given in supplementary material (Fig.~S5).
The uniformity concerning \ce{Ba, Ca, Ti} and \ce{Zr} species is similar to the \ce{BCTZ50} reference sample.
Given the very small amount of \ce{Ce} used, (maximum Ce molar concentration of 0.2\%), WDS could not be used to confirm the effective concentrations.
XPS Ce 3d spectrum have been acquired along the nominal gradient and are presented in Fig.~S6. 
An evolution of Ce 3d signal is evidenced along the gradient as expected. 
However quantification from the weak Ce 3d signal is not possible. 
One can note a very weak but non-zero signal at the nominally undoped position.
An average thickness of \qty{476}{nm} is obtained from WDS for the doped sample not considering its outskirt.

\begin{figure*}
    \centering
    \includegraphics[width=0.475\textwidth]{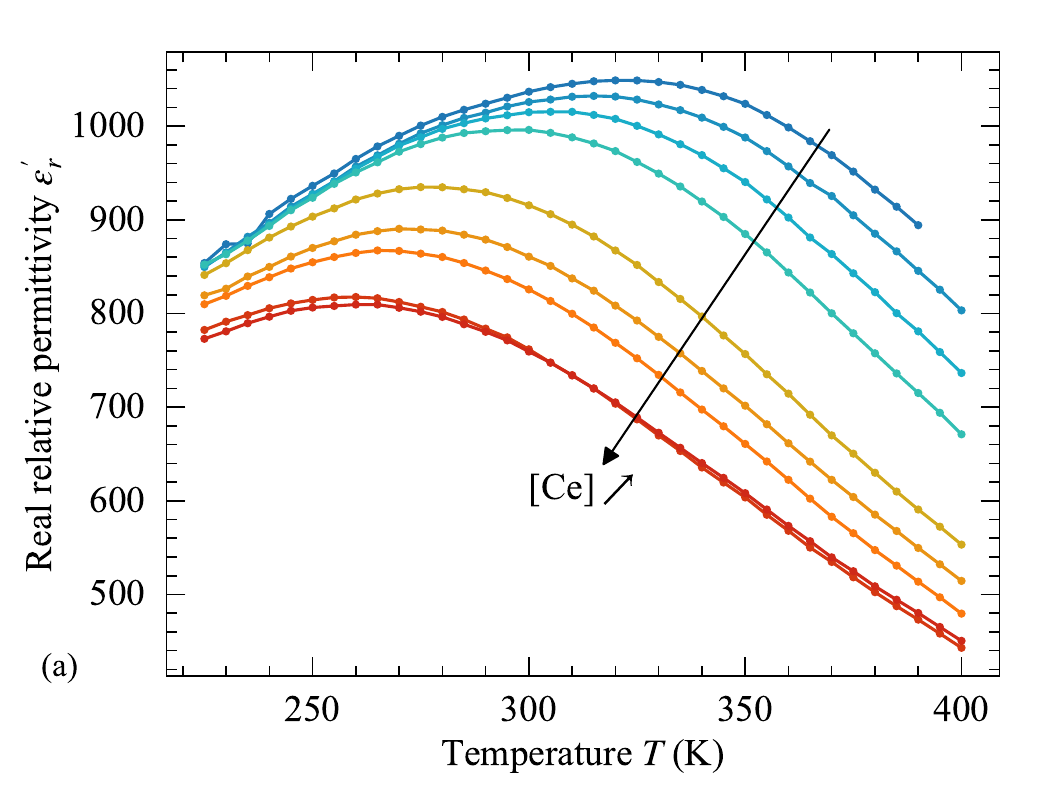}%
    \subfloat{\label{subfig:NJ406 realPerm TEMP}}%
    \hfill
    \includegraphics[width=0.475\textwidth]{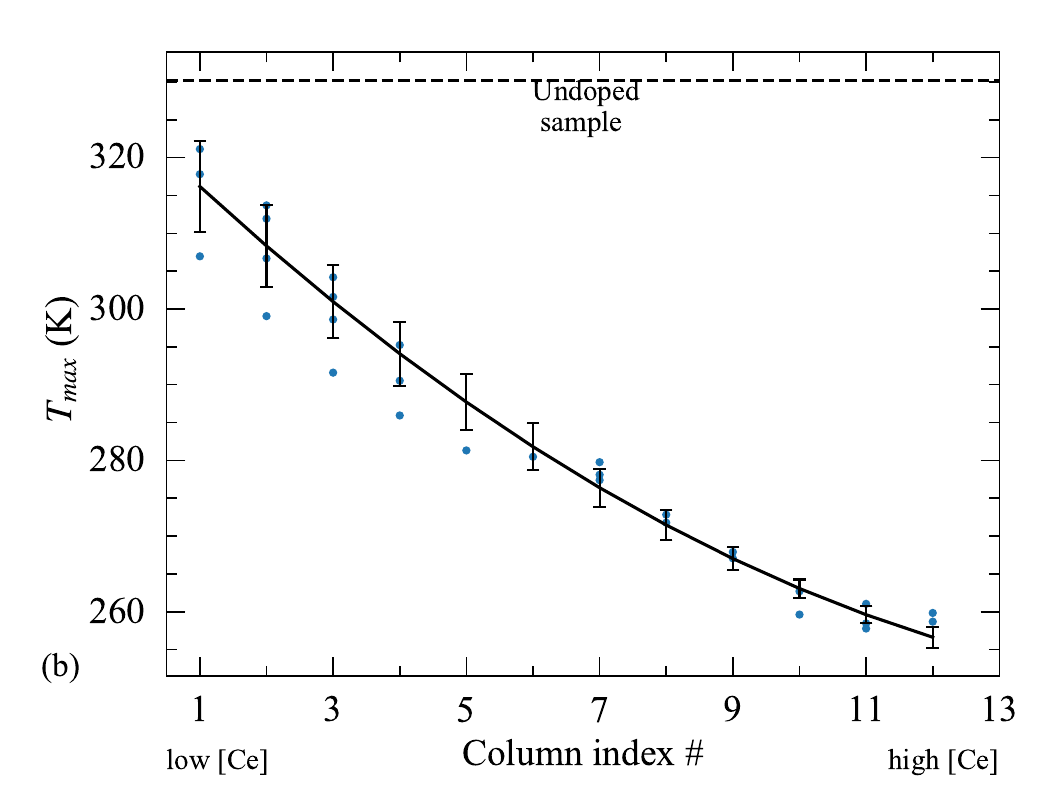}\\
    \subfloat{\label{subfig:NJ406 realPermTmax data}}%
    \includegraphics[width=0.522\textwidth]{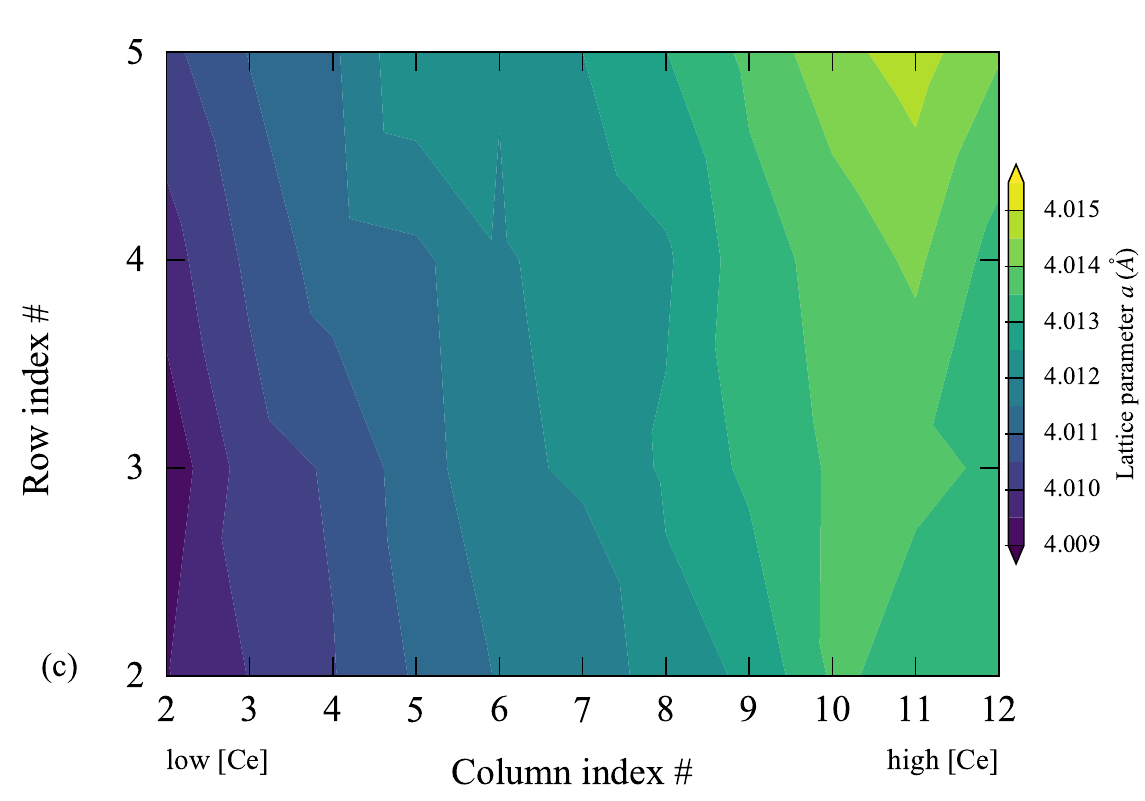}%
    \subfloat{\label{subfig:NJ406 ad222 COL}}
    \hfill
    \includegraphics[width=0.475\textwidth]{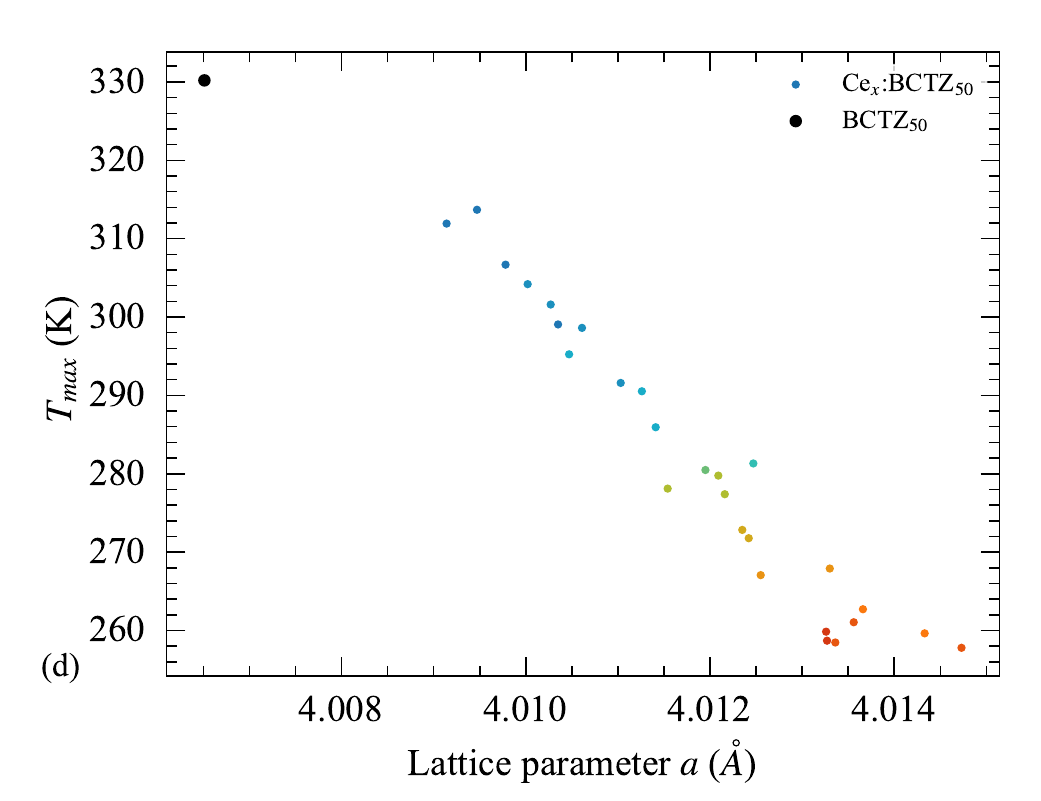}%
    \subfloat{\label{subfig:NJ406 ad222 realPermTmax}}%
    \caption{Relative dielectric permittivity as a function of temperature for some selected capacitors having different column number (a). $T_{\mathit{max}}$ as a function of the column number (b). Lattice parameter computed using the interplanar spacing as a function of position on the sample (c). $T_{\mathit{max}}$ as a function of the $d_{204}$ (orthorhombic) or $d_{222}$ (tetragonal or cubic) lattice parameter (d). }
    \label{fig:NJ406 realPerm Tmax ad222}
\end{figure*}

Fig.~\ref{fig:NJ406 realPerm Tmax ad222} shows the effect of the Ce on the dielectric permittivity on some representative capacitors.
When the \ce{Ce} content increases, $T_{\mathit{max}}$ decreases from \qty{320}{K} to \qty{260}{K} and the maximum permittivity value goes from 1050 to 810.
Thus, at ambient temperature, for undoped and low Ce-content (columns 1/2) the material is into the ferroelectric phase. 
For a Ce-content estimated to \qty{0.02}{\%} (columns 3/4), the material is in the transition region between ferroelectric and paraelectric.
For larger Ce-content (above column 5), the material is in the paraelectric phase.
The decrease of $T_{\mathit{max}}$ suggests the Ce substitutes the A site element (Ba or Ca here) since for B site substitution, no change of the phase transition temperature is visible in \ce{BaTiO3} \cite{HwangJACS2001}.
Nevertheless, the change in $T_{\mathit{max}}$ in our case is 15 times larger \qty{-70}{K}/\qty{0.2}{mol\%} vs \qty{-4.5}{K}/\qty{0.2}{mol\%} in \cite{HwangJACS2001}.

Room temperature (RT) X-ray $\theta-2\theta$ micro-diffraction (\qty{500}{\um} collimator) patterns were acquired for capacitors from column 2 to 12 for lines 2 to 5 (44 patterns) around $2\theta = \qty{83.5}{\degree}$.
The beam footprint on the sample in the gradient direction is \qty{500}{\um} and perpendicular to this direction \qty{670}{\um} ($\simeq \frac{500}{\cos\theta}$).
All these patterns exhibit a single peak (see Fig. S2 in supplementary material) which could be indexed (222)$_{C,T}$ , (204)$_{O}$ or (240)$_{O}$, (006)$_{R}$ or (042)$_{R}$ depending on the actual symmetry (respectively cubic, tetragonal, orthorhombic, rhombohedral). 
When $T_{\mathit{max}}$ is below RT (columns 5-12) the RT symmetry is cubic (paraelectric phase) while for $T_{\mathit{max}} >$ RT we cannot conclude on the symmetry. 
The FWHM of the observed diffraction peaks is independent of Ce content. 
The interreticular distance was extracted versus sample position using Bragg relation. 
The corresponding pseudo-cubic lattice parameter is represented as a map in Fig.~\ref{subfig:NJ406 ad222 COL}. 
A predominantly horizontal gradient of the lattice parameter is observed and correlates to the nominal Ce gradient: the lattice parameter progressively increases when the Ce content increases, confirming that Ce is entering into the perovskite phase. 
Taking a closer look to Fig.~\ref{subfig:NJ406 ad222 COL} one can see that the lines are not strictly vertical, indicating a slight variation of the lattice parameter, and hence of the Ce content, along columns with nominal constant composition. 
In fact there is a quasi-continuum of lattice parameter (and hence of Ce doping) across the sample correlated to a quasi-continuum of $T_{\mathit{max}}$ as can be seen of Fig.~\ref{subfig:NJ406 ad222 realPermTmax}.

The highest observed $T_{\mathit{max}}$ for the \ce{Ce_{x}{:}BCTZ} sample ($\sim$\qty{322}{K} for column 1 line 2 see Fig.~\ref{subfig:NJ406 realPermTmax data}) is lower than the value measured for the \ce{BCTZ50} sample (\qty{330}{K}) while the smallest lattice parameter in the \ce{Ce_{x}{:}BCTZ} (\qty{4.009}{\angstrom}) is still higher than for \ce{BCTZ50} undoped  sample (\qty{4.006}{\angstrom}) even for a zero Ce nominal content. 
This is attributed to the presence of a very small amount of Ce even on the side expected to be undoped as can be seen from Ce 3d XPS signal (Fig.~S6). 
This is confirmed by the fact that the data point of the undoped sample (in black) in Fig.~\ref{subfig:NJ406 ad222 realPermTmax} follows the trend but is far away from the rest of data points from \ce{Ce_{x}{:}BCTZ} sample (colored)
\begin{figure}
    \centering
    \includegraphics[width=0.48\textwidth]{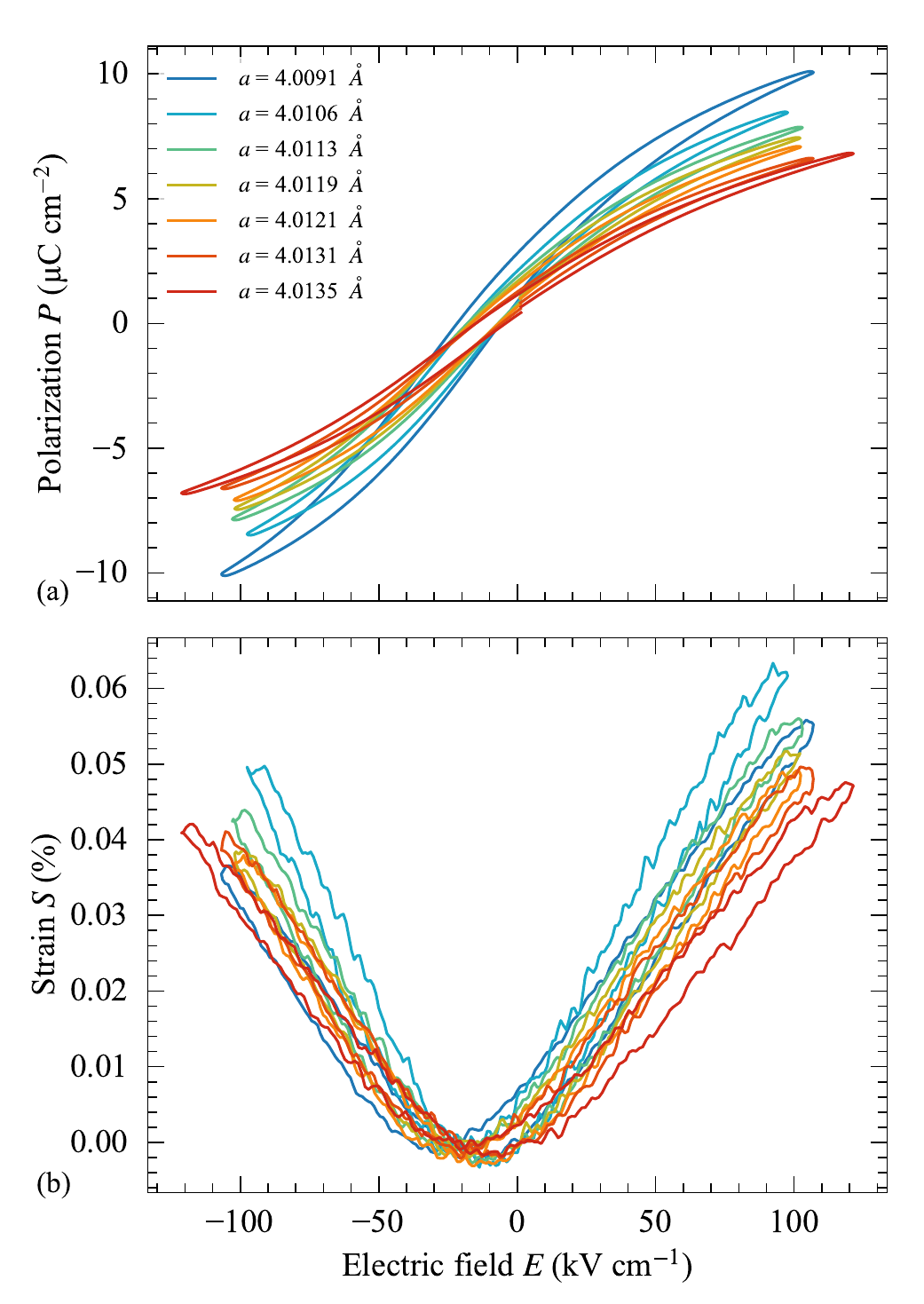}    
    \subfloat{\label{subfig:NJ406 polarization}}%
    \subfloat{\label{subfig:NJ406 strain}}%

    \includegraphics[width=0.48\linewidth]{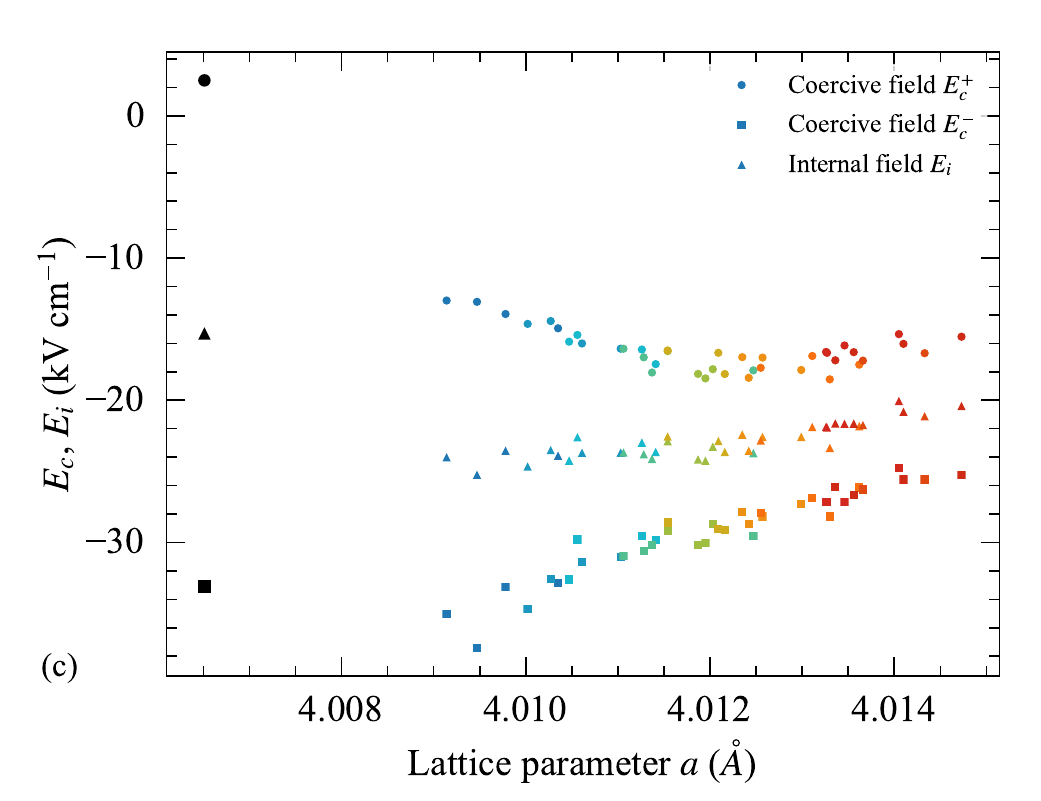}%
    \subfloat{\label{subfig:NJ406 coercive field}}%
    \caption{Polarization (a) and strain (b) as a function of the applied electric field for different compositions (having a different lattice parameter) present on the \ce{Ce_{x}{:}BCTZ50} sample. Coercive and internal field as a function of the lattice parameter (c). Black symbols correspond to the undoped sample.}
    \label{fig:NJ406 polarization strain}
\end{figure}

\begin{figure*}
    \centering
    \includegraphics[width=\textwidth]{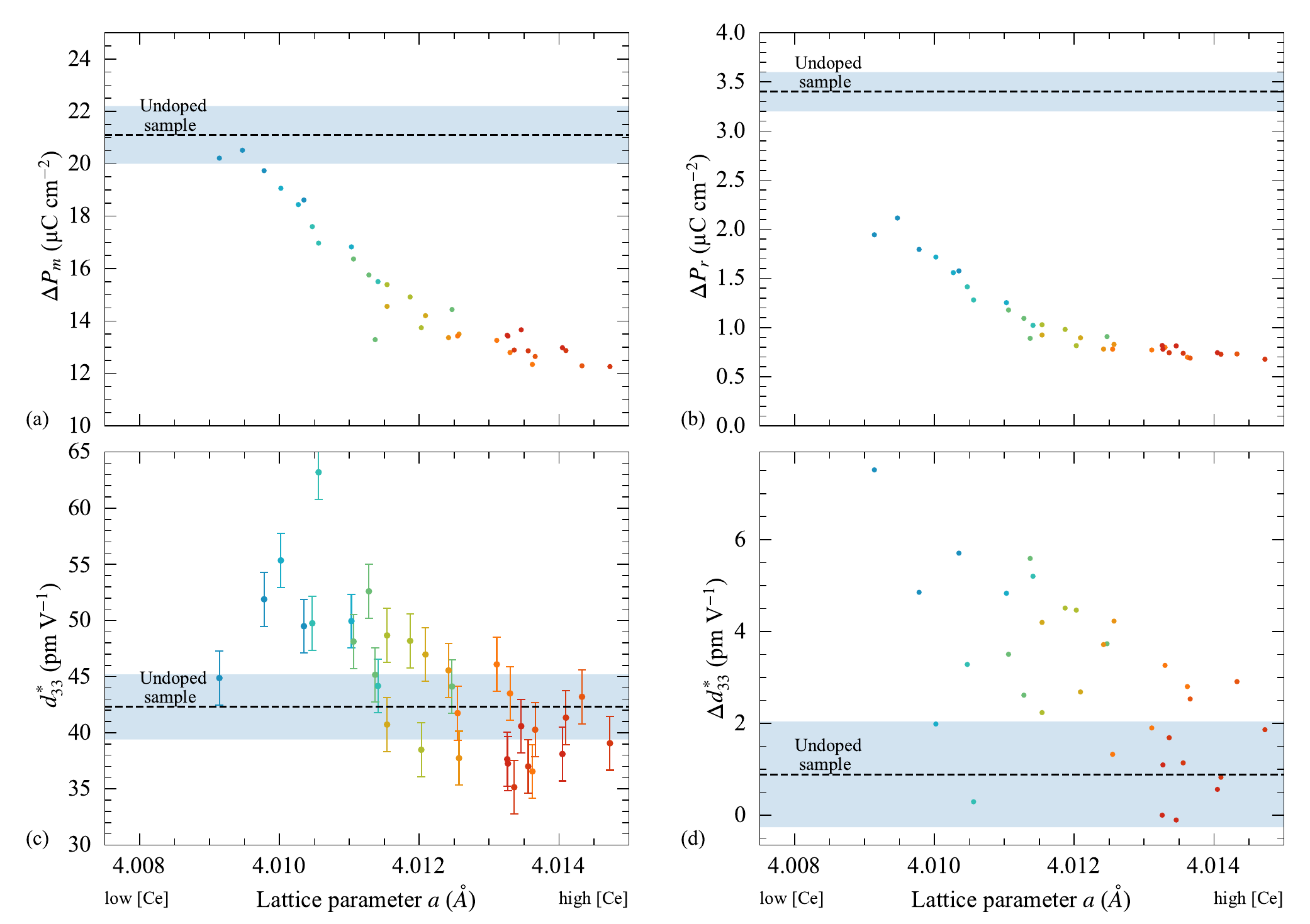}%
    \subfloat{\label{subfig:NJ406 2Pm data ad222}}%
    \subfloat{\label{subfig:NJ406 2Pr data ad222}}%
    \subfloat{\label{subfig:NJ406 d33 data ad222}}%
    \subfloat{\label{subfig:NJ406 delta d33 data ad222}}%
    \caption{Maximum polarization (a), remnant polarization (b), piezoelectric coefficient (c) and difference between positive and negative branch piezoelectric coefficient (d) for the \ce{Ce_{x}{:}BCTZ50} sample, as a function of the lattice parameter. }
    \label{fig:NJ406 2Pm 2Pr d33}
\end{figure*}

The evolution of the phase transition temperature $T_{\mathit{max}}$ is directly linked to the lattice parameter variation (Fig.~\ref{subfig:NJ406 ad222 realPermTmax}) as $T_{\mathit{max}}$ regularly decreases when the lattice parameter increases corresponding to the increase of Ce content. 
As a direct measurement of Ce content is not experimentally accessible given its small content, we will use the lattice parameter as an indirect indication of actual Ce content variation instead of the nominal Ce content in the following figures.

\cref{subfig:NJ406 polarization,subfig:NJ406 strain} show the polarization and the strain for several electrodes on the \ce{Ce_{x}{:}BCTZ50} sample.
When the Ce content increases, maximum polarization decreases and the $P(E)$ loop goes to a slimmer shape.
An horizontal shift of the $P(E)$ and $S(E)$ loops is present, similar to what has been mentioned for the undoped sample.
\cref{subfig:NJ406 coercive field} shows the coercive and the internal field as a function of the lattice parameter.
For low amount of Ce (lattice parameter between \qty{4.009}{\angstrom} and \qty{4.0106}{\angstrom}), the lower coercive field $E_{c}^{-}$ is similar to undoped BCTZ whereas the upper coercive field $E_{c}^{+}$ is more negative (\qty{-12}{\kV\per\cm} vs \qty{2.5}{\kV\per\cm}), resulting in a larger internal bias field (\qty{-24}{\kV\per\cm} vs \qty{-16}{\kV\per\cm}).
When the Ce content increases, the difference between coercive field decreases, corresponding to the slimmer shape described previously, and the internal bias field slightly decreases (in absolute value) from \qty{-24}{\kV\per\cm} to \qty{-20}{\kV\per\cm}.
This evolution of the internal bias may reveal an evolution of the PNR density with the amount of Ce\cite{NadaudAPL2024}.

Fig.~\ref{fig:NJ406 2Pm 2Pr d33} shows the ferroelectric and piezoelectric properties for the \ce{Ce_{x}{:}BCTZ50} sample.
When Ce content increases, maximum polarization decreases from $\Delta P_{m} = \qty{21.1(11)}{\micro\coulomb\per\cm\squared}$ to $\Delta P_{m} = \qty{13.4(3)}{\micro\coulomb\per\cm\squared}$ and the remnant polarization decreases from $\Delta P_{r} = \qty{3.4(2)}{\micro\coulomb\per\cm\squared}$ to $\Delta P_{r} = \qty{0.80(3)}{\micro\coulomb\per\cm\squared}$.
This large decrease of the polarization has been observed for BCTZ ceramics only for Ce content above \qty{0.25}{\%} \cite{HayatiJAC2019}.
For this reason, the decrease of maximum and remnant polarizations is attributed here to the decrease of the phase transition temperature $T_{\mathit{max}}$ when the Ce content increases: for composition having a lattice parameter above \qty{4.0105}{\angstrom}, ambient temperature is above $T_{\mathit{max}}$ and thus supposed to be in the paraelectric phase.

The piezoelectric coefficient sharply increases by \qty{50}{\%} from \qty{42.3(29)}{\pm\per\V} (for the undoped sample) to \qty{63(2.4)}{\pm\per\V} for a Ce-content estimated to about \qty{0.06}{\%} (lattice parameter of \qty{4.0105}{\angstrom}), then sharply decreases back to about \qty{42}{\pm\per\V} for a Ce-content estimated to \qty{0.11}{\%} (lattice parameter of \qty{4.0125}{\angstrom}). 
After that, a slow decrease of $d_{33}^{*}$ up to the maximum amount of cerium (\qty{0.2}{\%}, lattice parameter of \qty{4.0147}{\angstrom}).

The sharp $d_{33}^{*}$ peak could be evidenced solely because sub-nominal \qty{0.02}{\%} steps variations of Ce content were obtained in this quasi-continuous gradient sample. 
The very small amount of cerium needed to enhance the piezoelectric properties is close to what was found in ceramics\cite{HayatiJAC2019}. 
Indeed Hayati and co-workers showed that \qty{0.05}{\%} of Ce doping was better than \qty{0.025}{\%} and \qty{0.25}{\%}. 
However with such big steps they may have missed the optimum doping. 
This result strengthen the interest of the combinatorial approach and call for denser capacitors array associated with less steep gradient and thorough x-ray micro-diffraction analysis.

The difference between positive and negative fields piezoelectric coefficient $\Delta d_{33}^{*}$ is reported \cref{subfig:NJ406 delta d33 data ad222}.
A small amount of Ce (lattice parameter between \qty{4.009}{\angstrom} and \qty{4.0105}{\angstrom}) largely increases $\Delta d_{33}^{*}$ compared to the undoped sample, and when the amount of Ce increases, $\Delta d_{33}^{*}$ progressively decreases.
This evolution may be related to the evolution of the internal field (\cref{subfig:NJ406 coercive field}).

\subsection{Domain wall motion contribution}
To determine if the change in piezoelectric properties is intrinsic or extrinsic, the evolution of ferroelectric domain walls and their motions have been investigated by measuring the relative permittivity as a function of the AC field.
Due to the presence of irreversible domain wall motion contribution (also called pinning/unpinning), the relative permittivity increases when the AC measuring field increases\cite{taylorjap1997}.
In the case of homogeneous distribution of pinning centers, the relative permittivity linearly increases:
\begin{equation}
    \varepsilon_{r} = \varepsilon_{\mathit{r-l}} + \alpha_{r}E_{\mathit{AC}}
    \label{rayleigh}
\end{equation}
$\varepsilon_{\mathit{r-l}}$ correspond to the low field permittivity and $\alpha_{r}$ to the slope and represents the irreversible domain wall motion contribution.

In real materials, the distribution of pinning centers is not homogeneous and the linear increase is only visible after a given threshold\cite{HallF1999,SchenkPRA2018,TaylorAPL1998}.
For low field, domain walls can only vibrate around an equilibrium position and the contribution is called reversible.
In that case, the relative permittivity evolution with the measuring field can be described using the hyperbolic law:\cite{NadaudAPL2024,borderonapl2011,BaiCI2017}
\begin{equation}
    \varepsilon_{r} = \varepsilon_{\mathit{r-l}} + \sqrt{\varepsilon_{\mathit{r-rev}}^2 + (\alpha_{r}E_{\mathit{AC}})^2}
    \label{hyperbolic}
\end{equation}
with $\varepsilon_{\mathit{r-rev}}$ the reversible domain wall motion contribution, proportional to the domain wall density\cite{boserjap1987,NadaudAPL2022,BorderonSR2017}.

\begin{figure}
    \centering
    \includegraphics[width=0.48\textwidth]{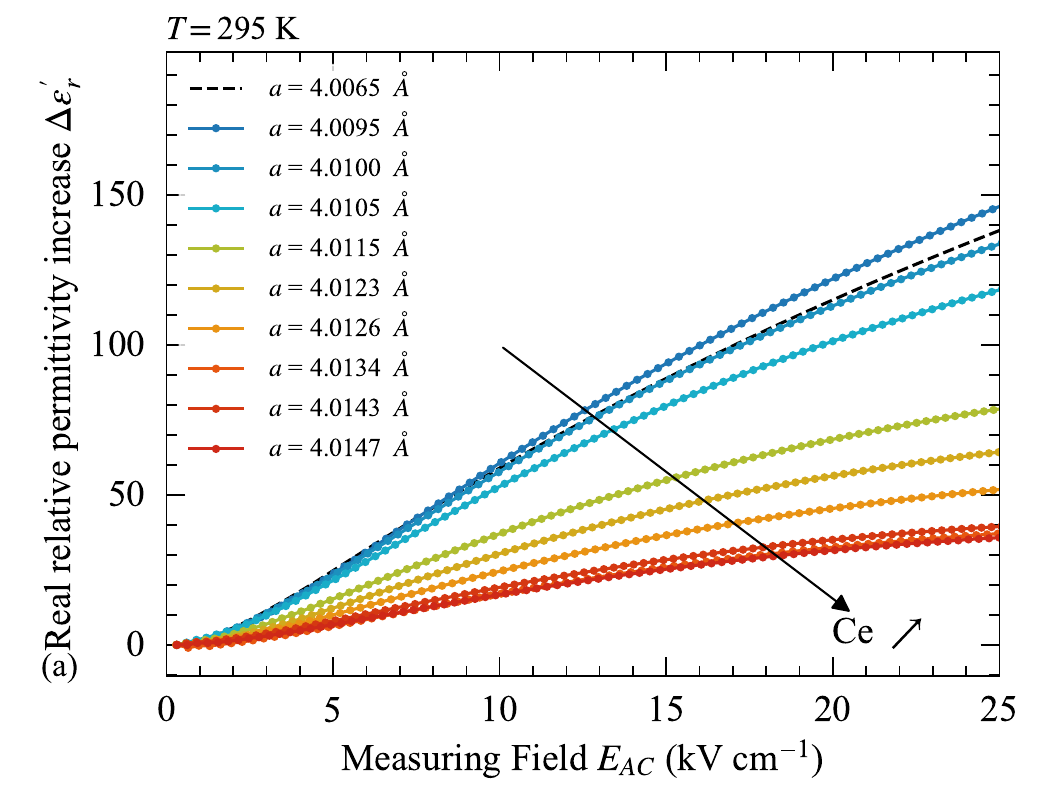}%
    \subfloat{\label{subfig:NJ406 realPermDelta OLEV}}%

    \includegraphics[width=0.48\textwidth]{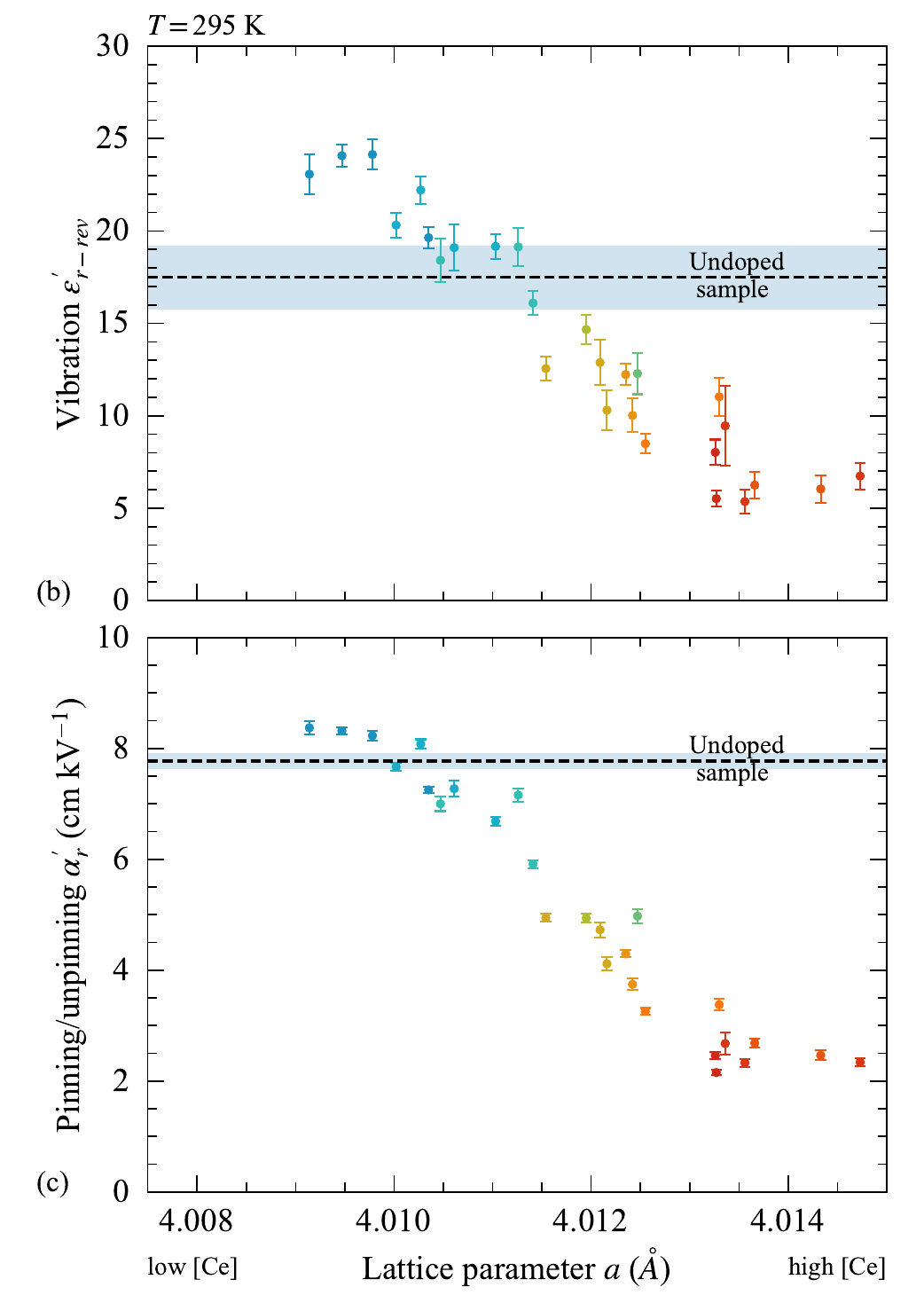}%
    \subfloat{\label{subfig:NJ406 Reve COL 295}}%
    \subfloat{\label{subfig:NJ406 Alph COL 295}}%
    \caption{Relative permittivity variation as a function of the AC measuring field for $T = \qty{295}{K}$ (a). Reversible (b) and irreversible (c) domain wall motion contributions to the permittivity as a function of the lattice parameter at $T = \qty{295}{K}$ for the \ce{Ce_{x}{:}BCTZ50} sample.  Black dashed lines correspond to values extracted for the \ce{BCTZ50} with, the shaded area represents the mean value and the standard deviation interval.}
    \label{fig:realPermDelta OLEV Reve Alph 295}
\end{figure}

Fig.~\ref{subfig:NJ406 realPermDelta OLEV} shows the relative permittivity variation (difference in relative permittivity with respect to the value for $V_{\mathit{AC}} = \qty{0.014}{V}$) as a function of the AC measuring field.
When cerium content increases, the relative permittivity variations decrease, which corresponds to a lower irreversible domain wall motion contribution.
By fitting the relative permittivity as a function of the AC measuring field using \eqref{hyperbolic}, it is possible to extract the reversible and irreversible domain wall motion contribution for different compositions and for different temperatures.

Fig.~\ref{subfig:NJ406 Reve COL 295},\subref*{subfig:NJ406 Alph COL 295} shows the reversible and irreversible domain wall motion contributions as function of the lattice parameter for $T = \qty{295}{K}$.
At room temperature (\qty{295}{K}), both reversible and irreversible domain wall motion contributions decrease when the Ce content increases and no sharp maximum is visible for a lattice parameter of \qty{4.0105}{\angstrom}, the electrode where the piezoelectric coefficient is maximum.
This indicates that the maximum value of the piezoelectric coefficient for this compound cannot be explained by a larger domain wall motion contribution (extrinsic).
The irreversible contributions for low Ce content (lattice parameter lower than \qty{4.011}{\angstrom}) of the \ce{Ce_{x}{:}BCTZ50} sample is very close to the ones obtained for the \ce{BCTZ50} (black dashed line) and the reversible contribution is slightly higher.

\begin{figure*}
    \centering
    \includegraphics[width=0.48\linewidth]{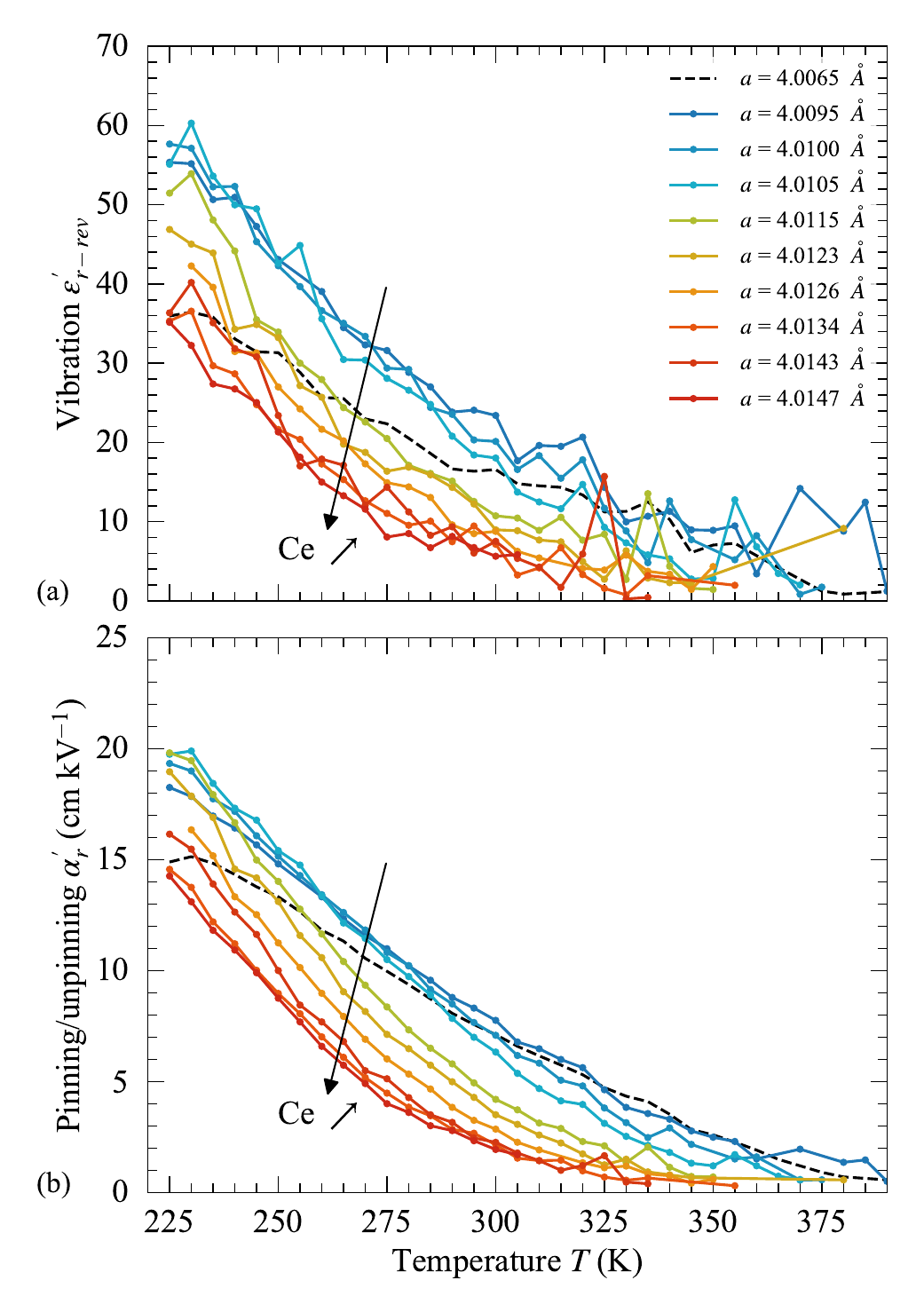}
    \subfloat{\label{subfig:NJ406 Reve COL TEMP}}%
    \subfloat{\label{subfig:NJ406 Alph COL TEMP}}%
    \includegraphics[width=0.48\linewidth]{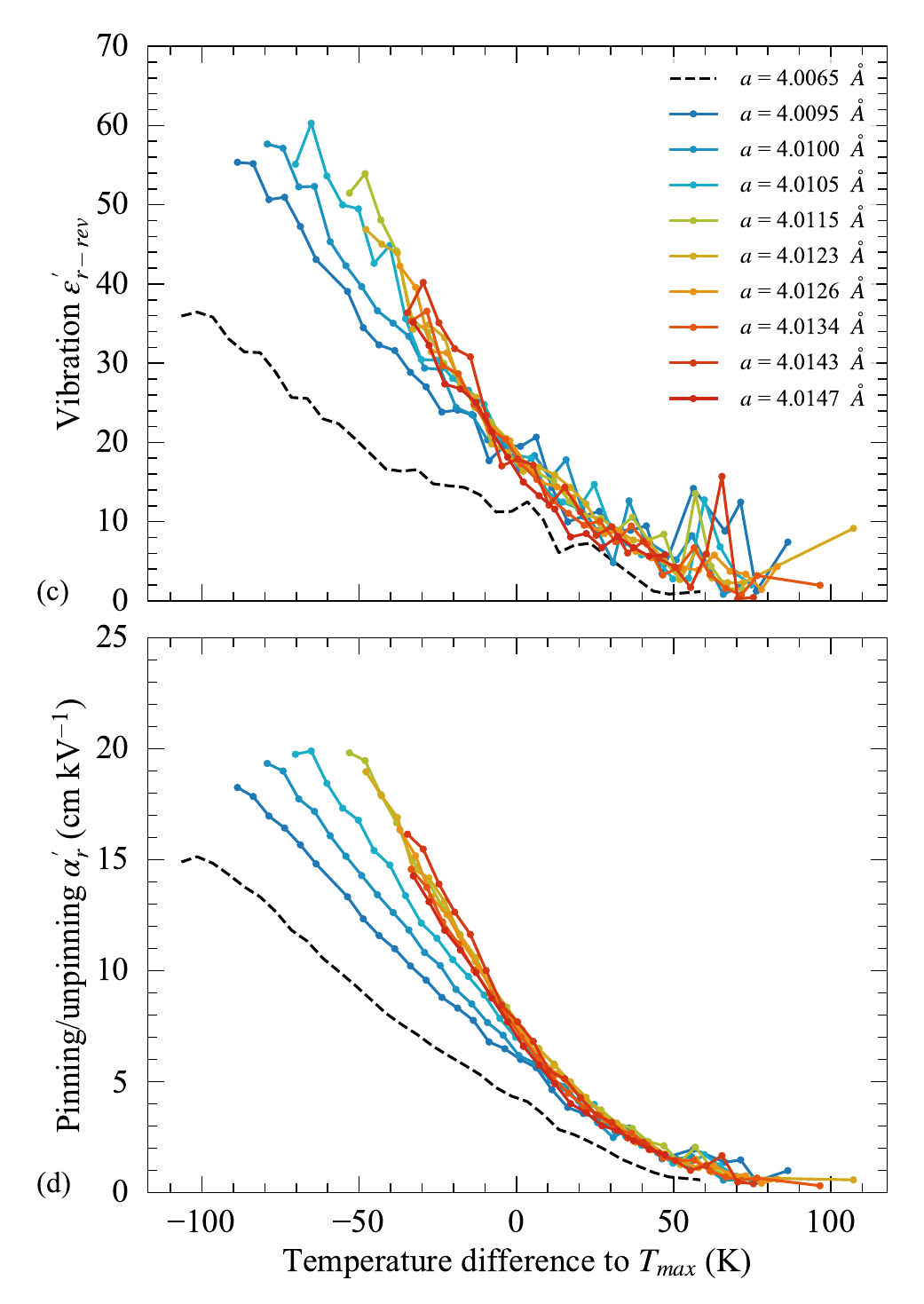}
    \subfloat{\label{subfig:NJ406 Reve COL TEMP offset}}%
    \subfloat{\label{subfig:NJ406 Alph COL TEMP offset}}%
    \caption{Reversible (a)-(c) and irreversible (b)-(d) domain wall motion contributions as a function of the absolute temperature (a)-(b) and the temperature relative to $T_{\mathit{max}}$ (c)-(d), for different compositions (having a different lattice parameter) present on the \ce{Ce_{x}{:}BCTZ50} sample. Black dashed lines correspond to values extracted for the \ce{BCTZ50}}
    \label{fig:}
\end{figure*}

Fig.~\ref{subfig:NJ406 Reve COL TEMP},\subref*{subfig:NJ406 Alph COL TEMP} present reversible and irreversible domain wall motion contributions as a function of temperature for the different compositions of the \ce{Ce_{x}{:}BCTZ50} sample.
For all compositions, reversible and irreversible domain wall motion contributions decrease when temperature increases, which is also the case for pure BCTZ\cite{NadaudAPL2024}.
This differs from what has been found for films of \ce{(Pb,Sr)TiO3}\cite{BaiCI2017} or \ce{0.5PbYb_{1/2}Nb_{1/2}-0.5PbTiO3}\cite{BassiriGharbJE2007} for which the domain wall motion contribution increases when the temperature approaches $T_{\mathit{max}}$.
For low temperatures (below \qty{250}{K}), the reversible and irreversible domain wall motion contributions increase (lattice parameter lower than \qty{4.0105}{\angstrom}), then both contributions decrease, when the cerium content increases.

To determine if the decrease of domain wall motion contributions is linked to the decrease of $T_{\mathit{max}}$, it is useful to plot the reversible and the irreversible domain wall motion contributions as a function of the temperature offset to $T_{\mathit{max}}$ (Fig.~\ref{subfig:NJ406 Reve COL TEMP offset},\subref*{subfig:NJ406 Alph COL TEMP offset}).
Above $T_{\mathit{max}}$, the domain wall motion coefficients still decrease but are not null, corresponding to a residual ferroelectricity\cite{GartenJAP2014,GartenJACS2016}, which persists here \qty{70}{\K} above $T_{\mathit{max}}$.
Above $T_{\mathit{max}}$, when rescaled with respect to the temperature difference to $T_{\mathit{max}}$, the decrease rate and the reversible and irreversible values are the same for all the compositions.
For temperatures below $T_{\mathit{max}}$, reversible and irreversible domain wall motion contributions slightly increase (lattice parameter lower than \qty{4.0115}{\angstrom}), then stabilize, when the Ce content increases.

In ferroelectric materials, the density of domain walls depends on the distance with respect to the transition temperature \cite{EverhardtPRL2019}. 
Here, if it were the only reason for variations in the number of domain walls, all the curves in Fig.~\ref{subfig:NJ406 Reve COL TEMP offset},\subref*{subfig:NJ406 Alph COL TEMP offset} would overlap. 
The fact that it does not highlights that the energy landscape in the film depends also on local chemistry and strain changes induced by cerium-doping. 
While overall domain wall motion contributions decrease with increasing cerium-content (Fig.~\ref{subfig:NJ406 Reve COL TEMP},\subref*{subfig:NJ406 Alph COL TEMP}), because of the decrease of the phase transition temperature $T_{\mathit{max}}$, at a given temperature difference to $T_{\mathit{max}}$ a low amount of cerium is actually facilitating reversible and irreversible domain wall motions (Fig.~\ref{subfig:NJ406 Reve COL TEMP offset},\subref*{subfig:NJ406 Alph COL TEMP offset}).
This is also visible for $T = \qty{225}{K}$, far below $T_{\mathit{max}}$ (Fig.~\ref{subfig:NJ406 Reve COL TEMP},\subref*{subfig:NJ406 Alph COL TEMP}), where a small amount of cerium increases domain wall motion contributions.

\section{Conclusion}

In this article, we use high-throughput experiment to investigate the cerium doping effect on the dielectric and piezoelectric properties of \ce{BCTZ50} thin film deposited by pulsed laser deposition.
The first part is dedicated to the measurement on a homogeneous sample.
For all elements, the ratio between $3\sigma$ dispersion and the mean value is below \qty{3}{\%} and for the thickness, the ratio is \qty{10}{\%} (whole sample) but it decreases to \qty{7}{\%} when discarding edges of the sample.
On this first sample, $T_{\mathit{max}} = \qty{330.2(13)}{K}$, $\Delta P_{m} = \qty{21.1(11)}{\micro\coulomb\per\cm\squared}$, $\Delta P_{r} = \qty{3.4(2)}{\micro\coulomb\per\cm\squared}$ and $d_{33}^{*}=\qty{42.3(29)}{\pm\per\V}$ are obtained for the sample.
All theses elements reveal the capability of the deposition process to get homogeneous composition and dielectric/piezoelectric properties.

A sample with a composition gradient from \ce{BCTZ50} to \ce{Ce_{02}{:}BCTZ50} (0.2 mol\%) has been elaborated.
Even if the exact distribution of cerium is too small to be quantified using WDS, its presence is confirmed using XPS and its gradient is confirmed by the simultaneous evolution of the lattice parameter and the $T_{\mathit{max}}$ which are largely affected.
The quasi-continuous variation of the lattice parameter, representative of a quasi-continuous variation of the Ce content, allows evidencing a sharp peak in the piezoelectric coefficient in a very narrow doping window.
The piezoelectric coefficient $d_{33}^{*}$ increases by \qty{50}{\%} from \qty{42.3(29)}{\pm\per\V} (undoped) to \qty{63(2.4)}{\pm\per\V} for \qty{0.06}{\%} Ce-mol.
This result reinforce the interest of the combinatorial approach and call for even finer phase diagrams exploration based on denser capacitors array, less steep nominal gradient and thorough x-ray micro-diffraction analysis.

Measurement as function of the AC field and Rayleigh analysis have been done to determine if change of piezoelectric properties may be induced by a higher extrinsic contributions (domain wall motion).
For low cerium content, the domain wall motion contributions stay almost constant and then decrease when the cerium content increases.
No correlation with the maximum of the $d_{33}^{*}$ variation is observed, indicating that the increase of the piezoelectric coefficient is not induced by changes in dynamics and densities of domain walls. 
The measurements as a function of temperature reveal the energy landscape or the density of domain wall in the film depends on the absolute temperature.

\section*{Supplementary Material}
Supplementary material includes XRD and WDS characterizations.

\section*{Data availability}
The data that support the findings of this study are available from the corresponding author upon reasonable request.

\section*{Acknowledgments}
This work has been performed with the means of the CERTeM (microelectronics technological research and development center) of French region Centre Val de Loire.
This work was funded through the project MAPS in the program ARD+ CERTeM 5.0 by the Région Centre Val de Loire co-funded by the European Union (ERC, DYNAMHEAT, N°101077402) and by the French State in the frame of the France 2030 program through the French Research Agency (ANR-22-PEXD-0018). 
Fruitful discussion with Jean-Louis Longuet from CEA-Le Ripault on WDS analysis are gladly acknowledged.
Views and opinions expressed are however those of the authors only and do not necessarily reflect those of the European Union or the European Research Council. 
Neither the European Union nor the granting authority can be held responsible for them.

\section*{Conflict of Interest}
The authors declare no competing financial interest.

\bibliography{biblio_ferro.bib}

\clearpage
\includegraphics[width=\linewidth]{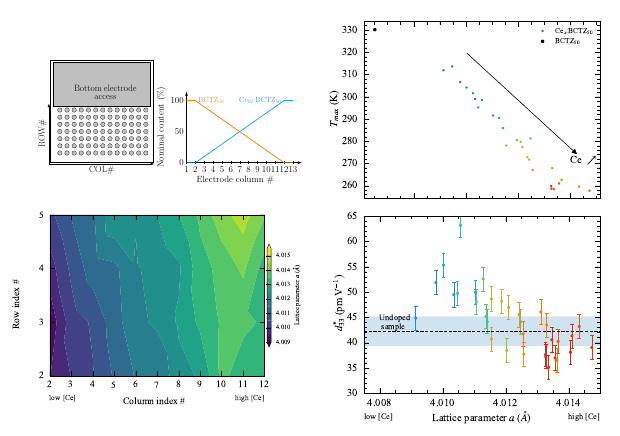}
For Table of Contents Only

\end{document}